\documentclass[aps,prb,twocolumn,showpacs,amsmath,amssymb]{revtex4-1}
\usepackage{graphicx}
\usepackage{dcolumn}
\usepackage{bm}

\usepackage{multirow}
\usepackage{epsfig}
\usepackage{amssymb}

\begin{document}
\title{Structural Tristability and Deep Dirac States in Bilayer Silicene on Ag(111) Surfaces}

\author{Zhi-Xin Guo}
\email[Present address: Department of Physics, Xiangtan University, Xiangtan, Hunan 411105, China]{}
\author{Atsushi Oshiyama}
\affiliation{Department of Applied Physics, The University of Tokyo, Hongo, Tokyo 113-8656, Japan}


\date{\today}

\begin{abstract}
We report on total-energy electronic-structure calculations in the density-functional theory performed for both monolayer and bilayer silicene on Ag(111) surfaces. The $\sqrt{3}\times\sqrt{3}$ structure observed experimentally and argued to be the monolayer silicene in the past [Chen {\it et al.}, Phys. Rev. Lett. {\bf 110}, 085504 (2013)] is identified as the bilayer silicene on the Ag(111) surface. The identification is based on our accurate density-functional calculations in which three approximations, the local density approximation, the generalized-gradient approximation, and the van-der-Waals-density-functional approximation, to the exchange-correlation energy have been carefully examined. We find that the structural tristability exists for the $\sqrt{3}\times\sqrt{3}$ bilayer silicene. The calculated energy barriers among the three stable structures are in the range of 7 - 9 meV per Si atom, indicating possible flip-flop motions among the three. We have found that the flip-flop motion between the two of the three structures produces the honeycomb structure in the STM images, whereas the motion among the three does the 1 $\times$ 1 structure. We have found that the electron states which effectively follow Dirac equation in the freestanding silicene couple with the substrate Ag orbitals due to the bond formation, and shift downwards {\em deep} in the valence bands. This feature is common to all the stable or metastable silicene layer on the Ag(111) substrate.
\end{abstract}


\maketitle

\section{Introduction}

A honeycomb-structured two-dimensional atomic layer consisting of group IV atoms exhibits peculiar electronic properties due to a fact that electrons near the Fermi level follow effectively the massless Dirac equation (Wyle equation).\cite{SW} A well-known and only unequivocally measured example is graphene where intriguing properties such as an anomalous quantum Hall effect is observed.\cite{geim,kim} Another element in group IV, Si, which has sustained our modern life, should exhibit such fascinating properties,\cite{takeda,cahangirov} thus opening a new door to the next-generation technology with its pronounced relativistic effects related to the spin degrees of freedom.\cite{yao,ezawa1,ezawa2}

However, no layered mother material such as graphite for graphene exists for Si in nature. Substrate materials are therefore indispensable to synthesize the honeycomb-structured Si atomic layers (silicene). Hence the identification of silicene-substrate interactions and their roles in the electronic properties of silicene becomes a central issue in silicene science.

Experimentally, Ag(111) surfaces are commonly used as such substrates and silicene layers on top are identified by scanning tunneling microscopy (STM) and spectroscopy (STS) measurements: The silicene mono-layers with the superperiodicities of 4$\times$4,\cite{vogt,lin,jamagotchian} $\sqrt{13}\times\sqrt{13}$,\cite{lin,jamagotchian} and 2$\sqrt{3} \times $ 2$\sqrt{3}$ \cite{jamagotchian,feng} with respect to 1 $\times$ 1 Ag(111) surface are observed, and the simulated STM images of theoretically determined structures reproduce the observed images excellently.\cite{guo1,guo2} As for the electron states near the Fermi level ($E_{\rm F}$), however, the situation is controversial: Angular-resolved photoelectron spectroscopy (ARPES) measurements \cite{vogt,avila} show the existence of the electron state with the linear energy dispersion (Dirac electron), whereas the density-functional calculations clarify the absence of Dirac electrons due to the strong silicene-substrate interactions;\cite{guo1,guo2,wang} it is of note that no Landau-level sequences peculiar to Dirac electrons have been observed.\cite{lin2}

The situation is further complicated for the newly found structure of silicene on Ag(111) with the periodicity of $\sqrt{3} \times \sqrt{3}$ with respect to the 1 $\times$ 1 silicene.\cite{chen1} From the interference patterns obtained by STS measurements, Chen {\it et al.} claimed the presence of Dirac electrons with the linear dispersion,\cite{chen1} whereas Arafune {\it et al.} deduced the contrary conclusion of the absence of Dirac electrons from essentially identical STS experiments.\cite{chen-arafune} Chen {\it et al.} \cite{chen2} also claimed the flip-flop motion between the two stable $\sqrt{3}$ structures to explain their hexagonal symmetry STM images at higher temperature and its freeze at lower temperature. Yet their evaluated energy barrier for the flip-flop motion, 38 meV per Si atom obtained by the density-functional calculations using VASP code,\cite{vasp1,vasp2} is too large to explain the frozen temperature 40 K. They also argued that the dispersion force, or van der Waals (vdW) force, is essential to obtain the rhombic $\sqrt{3}$ structure.\cite{chen2} However, this seems to be inconsistent with the formation of the covalent bonds for other stable structures.\cite{guo1,guo2} Recently, other experimental observations indicated that the existence of $\sqrt{3} \times \sqrt{3}$ monolayer (ML) silicene on Ag(111) is questionable, and suggested that Chen's observation should correspond to the bilayer (BL) silicene.\cite{Lay1,Lay2,Araf}

In this work, we focus on the $\sqrt{3} \times \sqrt{3}$ silicene on Ag(111) and identify the observed structure on the basis of the extensive density-functional calculations. Our results clearly contradict the existence of bistability between two rhombic $\sqrt{3}$ structures in ML silicene on Ag(111). Instead, in the BL silicene on the Ag(111) surface, we find the {\em tristability} among the three $\sqrt{3} \times \sqrt{3}$ structures. The calculated transition barriers are in the range of 7 to 9 meV per Si atom, inferring the possibility of the flip-flop motion at the low temperature. We also clarify that the vdW force is unimportant and the covalency plays a major role in the silicene-Ag interaction. This renders the Dirac states to be {\em deep} in the valence bands.

\section{Calculations}

The total-energy electronic-structure calculations have been performed in the density functional theory (DFT) using VASP code.\cite{vasp1,vasp2} The exchange-correlation functional, vdW-DF,\cite{dion,klimes} being capable of treating the dispersion force is adopted.\cite{ad1} To elucidate the role of the dispersion force (vdW force), we have also performed the calculations by the local density approximation (LDA) \cite{PZ} and the generalized gradient approximation (GGA).\cite{PBE} The electron-ion interaction is described by the projector augmented wave method,\cite{paw} and the cutoff energy of 250 eV in the plane-wave basis set is used. The Ag surface is simulated by a repeating slab model consisting of
a five-atomic-layer slab cleaved from the face-centered-cubic (fcc) Ag with the experimental lattice constant (4.09 {\AA}).
The Ag slab is separated from its images by the 14-{\AA}(17-{\AA}) vacuum region with single-layer(double-layer) Si atoms placed on top to simulate the ML(BL) silicene on the Ag(111) surface.
The geometry optimization is performed until the remaining forces become less than 0.02 eV/{\AA}. We have carefully examined the validity of the $k$-point mesh in Brillouin zone (BZ) integration, and found that the spacing between adjacent $k$-point being less than 0.016 \AA$^{-1}$ is sufficient to obtain converged results (the fine k-point mesh hereafter).

\section{Monolayer silicene on Ag(111)}
\begin{figure}
\begin{center}
\includegraphics[angle= 0, width=0.95\linewidth]{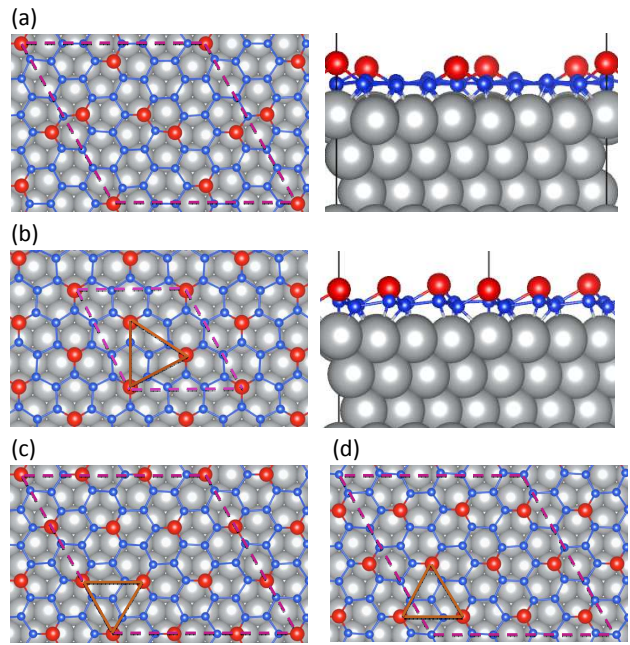}
\caption{
(color online)
Geometry optimized structures of monolayer (ML) silicene on the $7\times7$ Ag(111) surface (a) and on the $4\times4$ Ag(111) surface (b). In (a) and (b), the left and the right panels show the top and the side views, respectively. On the $7\times7$ Ag(111) and the $4\times4$ Ag(111), the ML silicene shows the superperiodicity of $3\sqrt{3}\times3\sqrt{3}$ and $\sqrt{3}\times\sqrt{3}$, respectively. The side views show the corrugation of the silicene layer, where the vertical unit cells are indicated by the solid (black) lines. Top views of the rhombic $\sqrt{3}\times\sqrt{3}$ structures obtained by using the insufficient $\Gamma$-point BZ sampling are shown in (c) and (d), which are mirror-symmetric to each other. The Si atoms in the top and bottom vertical positions are depicted by the large red and small blue balls, respectively. The large gray balls depict the positions of the substrate Ag atoms. The simulated lateral unit cells are indicated by the dashed (pink) lines, and the triangles denote the orientations of the protruded Si patterns with respect to Ag(111).
}
\label{strSL}
\end{center}
\end{figure}

We first consider the ML silicene with the periodicity of $3\sqrt{3}\times3\sqrt{3}$ on $7\times7$ Ag(111), which was reported to relax into two mirror-symmetric rhombic $\sqrt{3}\times\sqrt{3}$ structures by Chen {\it et al.}\cite{chen2}
After extensive search for stable geometries, we have reached a {\em single} $3\sqrt{3}\times3\sqrt{3}$ structure shown in Fig. \ref{strSL}(a). We were unable to obtain any $\sqrt{3}\times\sqrt{3}$ structure. We have found that the $\sqrt{3}\times\sqrt{3}$ structures obtained by Chen {\it et al.} [Figs. \ref{strSL}(c) and \ref{strSL}(d)] are reproduced only when we use the $\Gamma$-point BZ sampling in the calculations. In these $\sqrt{3}\times\sqrt{3}$ structures, one of the six Si atoms in a unit cell protrudes, whereas the remaining five Si keep nearly the same height in the bottom layer. Using the fine $k$-point mesh, we have found that these $\sqrt{3}\times\sqrt{3}$ structures obtained by the insufficient $k$-point sampling are unstable to further relaxation, and becomes the $3\sqrt{3}\times3\sqrt{3}$ structure shown in Fig. \ref{strSL}(a). In this stable $3\sqrt{3}\times3\sqrt{3}$ structure, 7 of the 54 Si atoms in a unit cell protrudes by about 1 {\AA} from the remaining 47 atoms [side view in Fig.~\ref{strSL}(a)].

\begin{table}
\caption{
Calculated cohesive energy $E_{\rm c}$ (eV/Si), binding energy $E_{\rm b}$ (eV/Si) and structural parameters for the stable $\sqrt{3}\times\sqrt{3}$ monolayer (ML1) and bilayer (BL1) silicene on Ag(111), as well as the $3\sqrt{3}\times3\sqrt{3}$ monolayer (ML2) silicene on Ag(111). $d_{\rm si-Ag}$ ({\AA}) is the spacing between the bottom silicene layer and the topmost Ag layer. Note that the silicene layer is buckled. In the monolayer case, $\Delta z_1$ ({\AA}) represents the average/maximum amount of the buckling of the silicene. In the bilayer case, $\Delta z_1$ and $\Delta z_2$ ({\AA}) represent the average/maximum amounts of the buckling of the lower- and upper-layer silicene, respectively, and $d_{\rm si-si}$ ({\AA}) is the interlayer distance defined as the minimum spacing between layers of the lower- and the upper-layer silicene.
}
\label{structure1}
\begin{tabular}{cccccccc}
  \hline
  \hline
             &~         &$E_{\rm c}$  &$E_{\rm b}$   & $d_{\rm si-Ag}$  & $d_{\rm si-si}$   & $\Delta z_1$  & $\Delta z_2$ \\
  \hline
             & vdW-DF   & 5.387      & 0.63        & 2.37              & ---        & 1.05/1.22     & ---   \\
  ML1        & GGA      & 5.20       & 0.43        & 2.37              & ---        & 1.07/1.23     & ---   \\
             & LDA      & 5.88       & 0.69        & 2.28              & ---        & 0.95/1.15     & ---    \\
                                            \\
             & vdW-DF   & 5.391      & 0.65        & 2.47              & ---        & 1.04/1.29     & ---   \\
  ML2        & GGA      & 5.24       & 0.45        & 2.47              & ---        & 1.06/1.34     & ---   \\
             & LDA      & 5.91       & 0.73        & 2.36              & ---        & 1.06/1.37     & ---    \\
                                             \\
            & vdW-DF   & 5.34       & 0.78        & 2.25              & 2.58        & 0.80/0.86     & 0.96/0.97  \\
  BL1       & GGA      & 5.18       & 0.57        & 2.27              & 2.58        & 0.81/0.86     & 0.98/0.98   \\
            & LDA      & 5.81       & 0.88        & 2.16              & 2.55        & 0.78/0.84     & 0.89/0.89   \\
            \\
  \hline
  \hline
\end{tabular}
\end{table}

We have also examined the possibility of obtaining the $\sqrt{3}\times\sqrt{3}$ ML silicene from other superperiodicity for which Chen {\it et al.} obtained the structure shown in Figs. \ref{strSL}(c) and \ref{strSL}(d) with the insufficient $\Gamma$-point sampling.\cite{chen2} We have examined the $3\times3$ silicene/$4\times4$ Ag(111), and found a {\em single} rhombic $\sqrt{3}\times\sqrt{3}$ structure shown in Fig.~\ref{strSL}(b). However, the orientation of the protruded Si pattern with respect to Ag(111) is obviously different from the mirror-symmetric structures obtained by Chen {et al.}. In addition, no other stable $\sqrt{3}\times\sqrt{3}$ structures which are related to Fig. \ref{strSL}(b) under the mirror symmetry operation of the silicene sheet exist. This indicates the importance of the silicene-substrate interaction.

The structural parameters for the stable $\sqrt{3}\times\sqrt{3}$ on $4\times4$ Ag(111) and the $3\sqrt{3}\times3\sqrt{3}$ on $7\times7$ Ag(111) are shown in Table.~\ref{structure1} along with the cohesive energy $E_{c}$ and the binding energy $E_{\rm b}$. They are defined as $E_{\rm c} = (E_{\rm Ag(111)} + N_{\rm Si} \mu_{\rm Si} - E_{\rm tot}) / N_{\rm Si}$ and $E_{\rm b} = (E_{\rm Ag(111)} + E_{\rm silicene} - E_{\rm tot}) / N_{\rm Si}$, respectively. Here $E_{\rm tot}$, $E_{\rm Ag(111)}$ and $E_{\rm silicene}$ are the total energies of the silicene on Ag(111), the clean Ag(111) surface and the freestanding silicene, respectively. $N_{\rm Si}$ is the number of Si atoms in a single layer of the silicene, and $\mu_{\rm Si}$ is the chemical potential of Si which is adopted as the total energy of an isolated Si atom.

The cohesive energy defined above is the energy gain to make silicene on the Ag(111) surface from the clean Ag surface plus the constituent Si atoms. The binding energy is the measure of the energy gain to put the free-standing silicene on the Ag(111) surface. In any case, $E_{\rm c}$ and $E_{\rm b}$ represent relative stability of various silicene layers on the Ag(111) surface.

Calculated cohesive energy for the stable $\sqrt{3}\times\sqrt{3}$ silicene on Ag(111) (5.387 eV/Si) is a bit smaller than that of the $3\sqrt{3}\times3\sqrt{3}$ on Ag(111) (5.391 eV/Si). Both of them are smaller than that of the most stable $3\times3$ silicene on Ag(111) (5.41 eV/Si in our calculation),\cite{guo1,guo2} which has been widely observed in experiments,\cite{vogt,lin} by about 20 meV/Si.
Moreover, the binding energies for both the stable $\sqrt{3}\times\sqrt{3}$ (0.63 eV/Si) and the $3\sqrt{3}\times3\sqrt{3}$ (0.65 eV/Si) structures are larger than the typical vdW interaction energy manifested in the cases of graphene on metal surfaces,\cite{Vanin} by an order of magnitude, showing the covalent-bonding interactions between silicene and Ag surface.

In the $3\sqrt{3}\times3\sqrt{3}$ and the stable $\sqrt{3}\times\sqrt{3}$ structures shown in Figs.~\ref{strSL}(a) and \ref{strSL}(b), the vertical spacing between the lower Si and the topmost Ag atoms, $d_{\rm si-Ag}$, is 2.47 and 2.37 {\AA}, respectively.
Considering that the atomic radii of Si and Ag are 1.18 {\AA} and 1.65 {\AA}, respectively, it is clear that the Si-Ag covalent bonds are formed. On the other hand, the vertical spacing between the protruded Si atoms and the topmost Ag atoms is around 3.5 {\AA}. This explains why the protruded Si atoms are located both on top and hollow sites of the Ag substrate in Figs.~\ref{strSL}(a) and \ref{strSL}(b).

We have also performed the LDA and GGA calculations, and obtained the essentially identical results with the vdW-DF calculations (Table. \ref{structure1}). This indicates that the dispersion (vdW) interactions only play a minor role in the ML silicene/Ag(111) system, in contradiction with the argument by Chen {\it et al.}.\cite{chen2} Our results obtained by the fine $k$-point mesh calculations conflict with those by Chen {\it et al.}, clarifying that the $\Gamma$-point sampling in the calculations is inadequate and leads to the incorrect results.

\section{Bilayer silicene on Ag(111)}

\begin{figure}
\begin{center}
\includegraphics[angle= 0,width=0.90\linewidth]{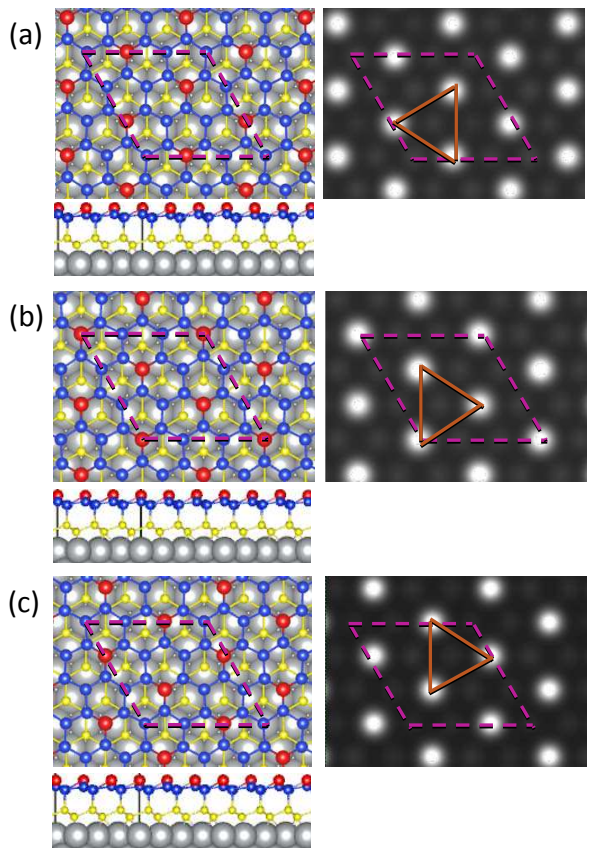}
\caption{
(color online)
Top and side views of the calculated stable structures (left panels) and corresponding STM images (right panels) of the three rhombic $\sqrt{3}\times\sqrt{3}$ bilayer (BL) silicene on Ag(111). The (a), (b) and (c) are the str1, str2 and str3 (see text), respectively. The large gray balls and the small yellow balls depict the positions of Ag atoms of substrate and the Si atoms of the lower-layer silicene, respectively. The large red balls and small blue balls depict the positions of protruded and unprotruded Si atoms of the upper-layer silicene. The simulated lateral unit cells in the top views are indicated by the dashed (pink) lines, and the vertical unit cells in the side views are indicated by the solid (black) lines. The obtained rhombic $\sqrt{3}\times\sqrt{3}$ periodicity is indicated by the solid (orange) lines.
}
\label{strBL}
\end{center}
\end{figure}

From our results obtained for the ML silicene described in the previous section,
we conclude that the honeycomb $\sqrt{3}\times\sqrt{3}$ silicene structure observed in the experiment does not correspond to the ML silicene. A plausible possibility is then the BL silicene.
Actually several experiments\cite{Lay1,Lay2,Araf} suggest the existence of bilayer and even a few-layer silicenes on the Ag(111) surface. We have then performed extensive search for the stable geometries of the BL silicene on Ag(111) with various commensurations of the two periodicities of the silicene and Ag(111). We have eventually found that the AB stacking $3\times3$ BL silicene on $4\times4$ Ag(111) is spontaneously relaxed into three $\sqrt{3}\times\sqrt{3}$ silicene structures, which we call str1, str2 and str3 hereafter, as shown in Fig.~\ref{strBL}.

Since the stacking of the two atomic layers of the BL silicene is not identified experimentally, we have also considered the case of
the AA stacking. Our calculations show that the AA stacking BL silicene is not relaxed to the $\sqrt{3}\times\sqrt{3}$ structure spontaneously. However, we have also found a stable AA stacking structure with the $3\times3$ periodicity. This shows that both the AA and AB bilayer silicene exist. We expect the experiments can also observe $3\times3$ AA stacking silicene on Ag(111) in the future.

\begin{figure}
\begin{center}
\includegraphics[angle= 0,width=0.9\linewidth]{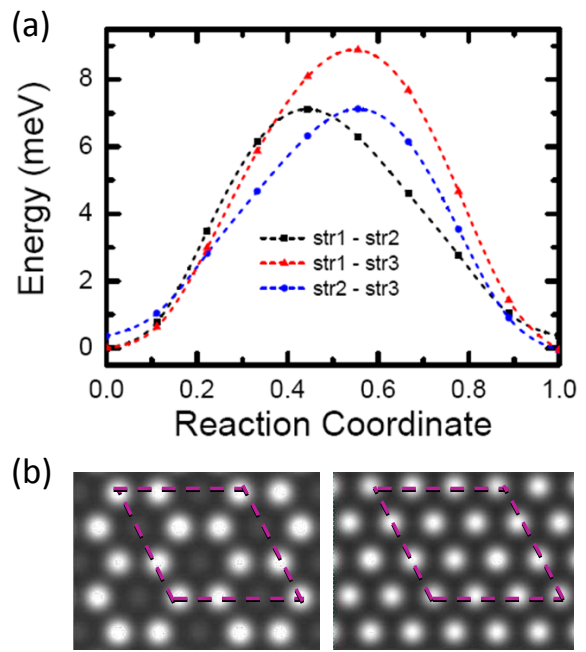}
\caption{
(color online)
(a) Calculated reaction energy profile among the three rhombic $\sqrt{3}\times\sqrt{3}$ structures in BL silicene on Ag(111), shown in Figs. \ref{strBL}(a), \ref{strBL}(b) and \ref{strBL}(c). The energy in the vertical axis is defined as $(E-E_{str1})/n$, where E and $E_{str1}$ are the total energy of the structure at reaction coordinate and the total energy of str1, respectively. $n$ is the number of atoms in the upper-layer silicene. The transition barrier between str1 and str2 (black dots), str2 and str3 (blue dots) are smaller than that between str1 and str3 (red dots) by about 2 meV/Si. (b) STM images obtained by superimposing the images of two (left panel) and three (right panel) stable structures.
}
\label{NEB}
\end{center}
\end{figure}

A common structural characteristic of the three stable $\sqrt{3}\times\sqrt{3}$ AB stacking structures is the protrusion of a single Si atom in the upper layer by 1 {\AA} from the remaining five Si atoms in the lateral $\sqrt{3}\times\sqrt{3}$ periodicity.
This is similar to the $\sqrt{3}\times\sqrt{3}$ structure of ML silicene in Fig. \ref{strSL}(b). In the lower-layer silicene, Si atoms are buckled with the amount of about 0.8 {\AA}, as they are on the (111) plane of diamond-structured Si.
The lowest-energy structure is str3. Other structures, str1 and str2, are higher in energy only by 0.06 meV and 0.36 meV, respectively, per Si atom of the upper-layer silicene, showing the existence of all three structures in experiments.
We have then identified the transition states among the three stable structures by the nudged elastic band (NEB) method.\cite{neb1,neb2} The obtained energy barrier for each reaction coordinate is shown in Fig. \ref{NEB}(a). Both the barriers from str1 to str2 and from str2 to str3 are 7.12 meV/Si, while it is a bit larger for the transition from str1 to str3 (8.88 meV/Si). It is noteworthy that these barriers are lower by an order than the values reported by Chen {\it et al.}\cite{chen2}

We have further calculated the STM images\cite{stm} of the three stable structures. As shown in the right panels of Fig.~\ref{strBL}, the STM images of the three structures we have found exhibit perfect rhombic $\sqrt{3}\times\sqrt{3}$ patterns. Each bright spot corresponds to a position of the protruded Si atom in the upper-layer silicene. These calculated images agree well with the STM image observed by Chen {\it et al.} [Fig. 1(e) of Ref.\onlinecite{chen2}] at low temperature (below 40K), where the big bright spots correspond to the protrusions of str2, whereas the small orange spots in the two different domains correspond to the protrusions of str1 and str3, respectively.

With increasing the temperature, the flip-flop motion between the two of the three stable structures may take place. In the case, the STM image should be the superposition of the two rhombic images. The left panel of Fig.~\ref{NEB}(b) shows such superposition of the two rhombic images, where the perfect honeycomb $\sqrt{3}\times\sqrt{3}$ image is prominent. The honeycomb image obtained agrees with the experimental images at higher temperature by Chen {\it et al}.\cite{chen1,chen2} Our results indicate that the honeycomb $\sqrt{3}\times\sqrt{3}$ silicene structure observed by the STM and identified as ML silicene is originated from the flip-flop motion of the rhombic $\sqrt{3}\times\sqrt{3}$ structures in {\em bilayer} silicene on Ag(111).

It should be mentioned that the transition barriers for the three rhombic structures are not large (below 9 meV/Si) so that the flip-flop motion is expected to happen frequently among the three at room temperature. In the case, the STM image of the $1\times1$ silicene structure, as calculated
in the right panel of Fig. \ref{NEB}(b), is also expected to be observed. Yet it has not been reported so far.

A common feature in the three $\sqrt{3}\times\sqrt{3}$ BL silicene structures is that all the protruded Si atoms in the upper-layer are located on the hollow sites of lower-layer silicene (left parts of Fig. \ref{strBL}), which makes them weakly interacted with the Si atoms in the
lower-layer silicene. Hence in contrast to the ML silicene case where the Si-Ag bond is reformed, the bond reformation is not required during the flip-flop motion among the three rhombic structures. This explains the substantially low energy barrier that we have obtained.

The structural parameters and corresponding cohesive and binding energies for the $\sqrt{3}\times\sqrt{3}$ BL silicene of str1 are shown in Table \ref{structure1}. The cohesive energy (5.34 eV/Si) is comparable with that of ML silicene and much larger than that of freestanding silicene (4.75 eV/Si in our calculation), showing the possibility of obtaining the $\sqrt{3}\times\sqrt{3}$ BL silicene in experiments.
The binding energy (0.78 eV/Si) between the BL silicene and Ag surface is even larger than that of the ML silicene on Ag(111). Here the binding energy is the energy gain in the deposition of BL silicene on Ag(111) surface, which is defined as $E_{\rm b} = (E_{\rm Ag(111)} + E_{\rm BL silicene} - E_{\rm tot}) / N^{\prime}_{\rm Si}$, where $N^{\prime}_{\rm Si}$ is the number of Si atoms in the lower-layer silicene. This implies the covalent bonding interactions with the substrates. The spacing between the lower-layer silicene and Ag surface (2.25 {\AA}) confirms the formation of Si-Ag bonds. The short silicene-silicene interlayer distance (2.58 {\AA}) also shows the formation of Si-Si bonds between layers.
The results for other two $\sqrt{3}\times\sqrt{3}$ BL silicene structures are nearly the same. We have also performed the LDA and GGA calculations and find that the results obtained through the different approximations are essentially identical (Table \ref{structure1}).

\begin{figure}
\begin{center}
\includegraphics[angle= 0,width=0.9\linewidth]{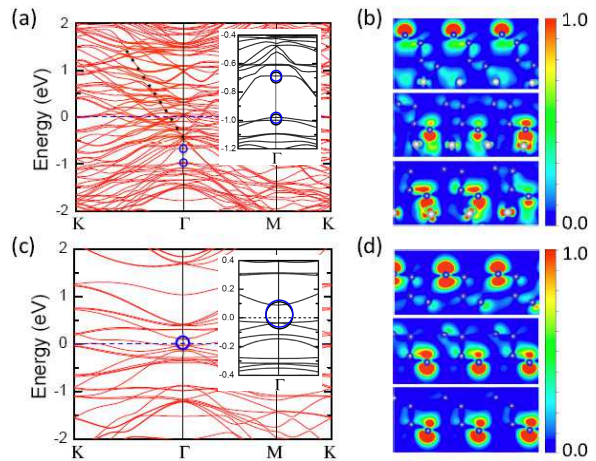}
\caption{
(color online)
Calculated energy bands (a) and contour plots of squared KS orbitals (b) of the $3\times3$ BL silicene/$4\times4$ Ag(111), where the silicene exhibits the $\sqrt{3}\times\sqrt{3}$ structure as shown in Fig. \ref{strBL}(a). The calculated energy bands (c) and contour plots of squared KS orbitals (d) of the freestanding $3\times3$ BL silicene peeled from the Ag(111) are also shown. The origin of the energy is set to be $E_{\rm F}$. The three states that have characters of the mixed $\pi$ ($\pi^{*}$) are indicated by the blue circles in (a) [(c)], and the corresponding KS orbitals are shown in (b) [(d)].
The insets of (a) and (c) show the enlarged figures for the energy bands around the blue circles for the silicene/Ag(111) and the peeled silicene, respectively. The linear band mainly consisting of the $s$ and $p$ orbitals of Ag is indicated by the black-dotted line. The gray and blue balls depict Ag and Si atoms, respectively.
}
\label{bdBL}
\end{center}
\end{figure}

We are now in a position to discuss the electronic structure. Fig. \ref{bdBL}(a) shows the calculated energy bands for the $3\times3$ BL silicene/$4\times4$ Ag(111), where the silicene presents the $\sqrt{3}\times\sqrt{3}$ periodicity [str1, shown in Fig. \ref{strBL}(a)]. Through the analyses of the Kohn-Sham (KS) orbitals, we have found no apparent energy band near $E_{\rm F}$ that shows the linear energy dispersion peculiar to Dirac $\pi$ and anti-bonding $\pi$ (labeled as $\pi^{*}$) states. Instead, we have found that the states with characters of $\pi$ and $\pi^{*}$ appear at 0.6-0.9 eV below $E_{\rm F}$ [Fig.~\ref{bdBL}(b)].

We have then calculated the energy bands for the freestanding $3\times3$ BL silicene which is peeled from the Ag surface [Fig. \ref{bdBL}(c)].
This freestanding BL silicene has the identical structure with the BL silicene in str1, except that the Si atoms in the lower layer lose their partner Ag atoms. Even for this freestanding BL silicene, the Dirac $\pi$ ($\pi^{*}$) states which show the linear dispersion are absent. In this case the energy bands which have characters of $\pi$ and $\pi^{*}$ appear near $E_{\rm F}$, with a band gap of 0.1 eV. The corresponding KS orbitals of these states are shown in Fig.~\ref{bdBL}(d), where obvious interlayer $\pi$-$\pi$ orbital mixing induced by the interlayer interactions as well as the intralayer $\pi$-$\sigma$ orbital mixing induced by the buckling of silicene layers are observed. Such mixing is the reason of the gap opening.

The $\pi$ and $\pi^{*}$ states shift upward near $E_{\rm F}$ from the deep position in the valence bands by peeling the BL silicene from the Ag substrate. The amount of the charge transfer between the Ag and the silicene is small, as is clarified in other stable silicene structures on the Ag(111).\cite{guo1,guo2} Hence it is now evident that the hybridization between the Si orbitals on the lower-layer silicene and the Ag orbitals on the substrate makes the $\pi$ and $\pi^{*}$ states deep in the valence bands and that cutting the bonds makes those states shift upward near $E_{\rm F}$.
This feature is common to all the silicene on the Ag(111) surfaces so far available.\cite{guo1,guo2}

Finally, we come to discuss the linear-like energy dispersion derived from the quasi-particle interference patterns in the experiments.\cite{chen1,chen2}
The Fermi velocity obtained from the linear-like energy dispersion [around $1.0 \times 10^{6} {\rm ms}^{-1}$] \cite{chen1,chen2} is two times larger than that of freestanding ML silicene (around $0.5 \times 10^{6} {\rm ms}^{-1}$)\cite{drummond}, suggesting that the experimental observation does not correspond to the Dirac states of silicene. Through the detailed analysis of KS orbitals for all the energy bands along the $\Gamma$-K direction in BZ, we have indeed found such a linear-like energy band, which appears in the experimental energy region (0.4 - 1.6 eV) [Fig.~\ref{bdBL}(a)] with the value of $(\partial \varepsilon / \partial k ) \hbar^{-1} = 1.04 \times 10^6 {\rm ms}^{-1}$.
From the KS-orbital analysis, the energy band mainly consists of the $s$ and $p$ orbitals of Ag
atoms located at subsurface atomic layers. This strongly indicates that the linear energy dispersion reported in the experiments corresponds to the $sp$ band of Ag substrate.

\section{Conclusion}

We have performed the density-functional calculations for both monolayer and bilayer silicene on Ag(111) surfaces. The $\sqrt{3}\times\sqrt{3}$ structure observed experimentally and argued to be the monolayer silicene in the past \cite{chen1,chen2} has been unequivocally identified as the bilayer silicene on the Ag(111) surface. The identification is based on our accurate density-functional calculations in which several approximations to the exchange-correlation energy have been carefully examined. We have found that the structural tristability exists for the $\sqrt{3}\times\sqrt{3}$ bilayer silicene. The calculated energy barriers among the three stable structures are in the range of 7 - 9 meV per Si atom, indicating possible flip-flop motions among the three. We have found that the flip-flop motion between the two of the three structures produces the honeycomb structure in the STM images, whereas the motion among the three does the 1 $\times$ 1 structure. We have found that the electron states which effectively follow Dirac equation in the freestanding silicene couple with the substrate Ag orbitals due to the bond formation, and shift downwards {\em deep} in the valence bands. This feature is common to all the stable or metastable silicene layer on the Ag(111) substrate.

\begin{acknowledgments}
This work was supported by the research project ``Materials Design through Computics" (http://computics-material.jp/index-e.html) by MEXT and also by ``Computational Materials Science Initiative" by MEXT, Japan.  Computations were performed mainly at Supercomputer Center in ISSP, University of Tokyo. ZX acknowledges the support of NSFC (Grant No. 11204259).
\end{acknowledgments}


\begin{thebibliography}{}
\bibitem{ad1}
The optB86b-vdW functional in Ref.\onlinecite{klimes} is adopted in vdW-DF calculations. The calculated lattice constant for the face-centered-cubic Ag is 4.10 {\AA}, which agrees well with experimental value (4.09 {\AA}).
\end{thebibliography}
\end{document}